\documentclass[fleqn,twoside]{article}
\usepackage{espcrc2}

\usepackage{graphicx}
\usepackage[figuresright]{rotating}

\newcommand{\AmS}{{\protect\the\textfont2
  A\kern-.1667em\lower.5ex\hbox{M}\kern-.125emS}}

\title{CDMS, Supersymmetry and Extra Dimensions}

\author{D.S.~Akerib\address[case]{Department of Physics, Case Western Reserve University, Cleveland, OH  44106, USA}, 
        M.J.~Attisha\address[brown]{Department of Physics, Brown University, Providence, RI 02912, USA}, 
        C.N.~Bailey\addressmark[case], 
        L. Baudis\address[aachen]{Department of Physics, RWTH Aachen University, Aachen, 52074, Germany }
						\thanks{Corresponding author, laura.baudis@rwth-aachen.de},
        D.A.~Bauer\address[fermi]{Fermi National Accelerator Laboratory, Batavia, IL 60510, USA},
        P.L.~Brink\address[stanford]{Department of Physics, Stanford University, Stanford, CA 94305, USA},
        P.P.~Brusov\addressmark[case], 
        R.~Bunker\address[barbara]{Department of Physics, University of California, Santa Barbara, CA 93106, USA},
        B.~Cabrera\addressmark[stanford], 
        D.O.~Caldwell\addressmark[barbara],
        C.L.~Chang\addressmark[stanford],
        J.~Cooley\addressmark[stanford], 
        M.B.~Crisler\addressmark[fermi],
        P.~Cushman\address[minnesota]{School of Physics \& Astronomy, University of Minnesota, Minneapolis, MN 55455, USA},
        M.~Daal\address[berkeley]{Department of Physics, University of California, Berkeley, CA 94720, USA}, 
        R.~Dixon\addressmark[fermi], 
        M.R.~Dragowsky\addressmark[case], 
        D.D.~Driscoll\addressmark[case],
        L.~Duong\addressmark[minnesota], 
        R.~Ferril\addressmark[barbara],
        J.~Filippini\addressmark[berkeley], 
        R.J.~Gaitskell\addressmark[brown],
        S.R.~Golwala\address[caltech]{Department of Physics, California Institute of Technology, Pasadena, CA 91125, USA}, 
        D.R.~Grant\addressmark[case],
        R.~Hennings-Yeomans\addressmark[case], 
        D.~Holmgren\addressmark[fermi],
        M.E.~Huber\address[denver]{Department of Physics, University of Colorado at Denver, Denver, CO~80217, USA}, 
        S.~Kamat\addressmark[case], 
        S.~Leclercq\address[florida]{Department of Physics, University of Florida, Gainesville, FL 32611, USA},
        A.~Lu\addressmark[berkeley],
        R.~Mahapatra\addressmark[barbara],
        V.~Mandic\addressmark[berkeley], 
        P.~Meunier\addressmark[berkeley],
        N.~Mirabolfathi\addressmark[berkeley],
        H.~Nelson\addressmark[barbara], 
        R.~Nelson\addressmark[barbara], 
        R.W.~Ogburn\addressmark[stanford], 
        T.A.~Perera\addressmark[case], 
        M.~Pyle\addressmark[stanford],
        E.~Ramberg\addressmark[fermi], 
        W.~Rau\addressmark[berkeley], 
        A.~Reisetter\addressmark[minnesota],
        R.R.~Ross\address[lawrence]{Lawrence Berkeley National Laboratory, Berkeley, CA 94720, USA}\addressmark[berkeley],
        T.~Saab\addressmark[florida],
        B.~Sadoulet\addressmark[lawrence]\addressmark[berkeley],
        J.~Sander\addressmark[barbara],
        C.~Savage\addressmark[barbara],
        R.W.~Schnee\addressmark[case],
        D.N.~Seitz\addressmark[berkeley],
        B.~Serfass\addressmark[berkeley], 
        K.M.~Sundqvist\addressmark[berkeley],
        J-P.F.~Thompson\addressmark[brown],
        G.~Wang\addressmark[case]\addressmark[caltech],
        S.~Yellin\addressmark[stanford]\addressmark[barbara], 
        J.~Yoo\addressmark[fermi] and
        B.A.~Young\address[clara]{Department of Physics, Santa Clara University, Santa Clara, CA 95053, USA} }

\begin{document}

\begin{abstract}

The CDMS experiment aims to directly detect massive, cold dark matter particles originating from the Milky Way halo. 
Charge  and lattice excitations are detected after a particle scatters in a Ge or Si crystal 
kept at $\sim$30\,mK, allowing to separate nuclear recoils from the dominating electromagnetic background.
The operation of 12 detectors in the Soudan mine for 75 live days in 2004 delivered 
no evidence for a signal, yielding stringent limits on dark matter candidates from supersymmetry and universal 
extra dimensions. Thirty Ge and Si detectors are presently installed in the Soudan cryostat, and operating at base 
temperature. The run scheduled to start in 2006 is expected to yield a one order of magnitude increase in dark matter 
sensitivity.

\end{abstract}

\maketitle

\section{The CDMS Experiment at Soudan}

The Cryogenic Dark Matter Search (CDMS) experiment is designed to search for cold dark matter in form of weakly 
interacting massive particles (WIMPs). These hypothetical thermal relics of the big bang are expected to 
have masses and annihilation cross sections at the weak scale if their relic density is around around $\Omega$=0.2.
Examples are the lightest supersymmetric particle (LSP, or neutralino) and the lightest Kaluza-Klein 
particle (LKP) in theories with large universal extra dimensions. 

In CDMS, WIMPs can be detected via their elastic scattering with Ge or Si nuclei. Events with nuclear recoil energies of a 
few tens of keV and rates below 1~event kg$^{-1}$d$^{-1}$ are expected.
The experiment is operated at the Soudan Underground Laboratory in Minnesota (at 2080 m.w.e.), in a dedicated 
low-background facility. At its core are thirty z-dependent ionization- 
and phonon-mediated (ZIP) detectors, arranged in 5 towers. Within one tower the ZIPs are 
stacked 2\,mm apart, with no intervening material. Such close packing shields the crystals against surface 
low-energy electrons, and allows to identify multiply scattered events. 
A particle interaction in a ZIP detector deposits energy into the 
crystal via lattice vibrations, or phonons, and via charge excitations, or electron-hole pairs. The simultaneous 
measurement of both phonon and ionization signals allows to accurately measure the recoil energy, and to distinguish 
between nuclear- and electron recoil events. This distinction becomes possible since nuclear recoils produce fewer 
charge pairs than electron recoils of the same energy. The ionization yield, defined as the ratio of 
ionization to full recoil energy, is unity for electron recoils with complete charge collection, 
and $\sim$0.3 ($\sim$0.25) for nuclear recoils in Ge (Si) with recoil energies above 20\,keV.

\begin{figure}[h]
\hspace*{-0.5cm}{\includegraphics[width=.5\textwidth]{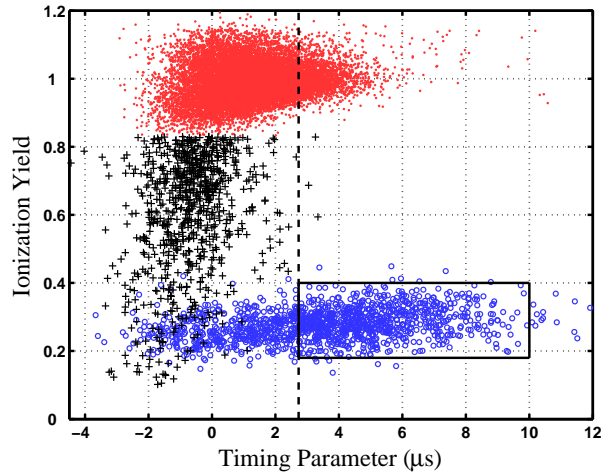}} 
\caption{Ionization yield versus phonon timing parameter for $^{133}$Ba gamma calibration events (dots and crosses)
and $^{252}$Cf neutron calibration events (circles). Low-yield $^{133}$Ba events (crosses) 
have small values of the timing parameter, a typical cut is indicated by the dashed vertical line.} 
\label{fig1a}
\end{figure}

ZIP detectors measure the athermal phonos created in a particle interaction using quasiparticle-assisted
electrothermal-feedback transition-edge sensors (QETs) photolitographically patterned onto 
one of the crystal surfaces. The QETs, which are made of 1\,$\mu$m wide strips of tungsten connected to eight 
superconducting aluminum fins are divided into four independent channels, each consisting of 1036 QETs 
operated in parallel. The narrow tunsten strips form the transition-edge sensors (TESs), which are voltage biased 
and kept stably within their superconducting to normal transition by electrothermal feedback based on Joule self-heating.
Energy deposited in the tungsten electron system raises the temperature of the film, increasing its resistance. 
The corresponding change in current is detected by a high-bandwidth SQUID array.
The phonon pulse rise times are sensitive to the phonon arrival times at each of the four quadrants and allow 
to localize an event in the x-y plane. In addition, events occurring near the detector's surface display faster 
rise times than bulk events, allowing for a fiducial volume cut.
For the ionization measurement, a drift field of a few V/cm is applied across the crystal using 
electrodes deposited on the two faces of each detector. The electrodes are segmented radially,
in an outer annular guard ring and a central disk-shaped volume, a design optimized to veto events 
near the edge of the crystals. A detailed description of the CDMS apparatus and shield is given 
in \cite{cdms_prd}.

\section{CDMS data, results and current status}

The most recent CDMS data were collected between March and August 2004
with two towers (6 Ge and 6 Si ZIPs)  for a total of 74.5 live days yielding  
an exposure of 34\,kg\,d in Ge and 12\,kg\,d in Si in the 10-100\,keV nuclear recoil energy range \cite{cdms_si}. 
To avoid bias, the analysis was perfomed blind, whereby events in and near the signal region were 
masked in the WIMP search data sets. The cuts defining a signal were determined using calibration 
data from $^{133}$Ba and $^{252}$Cf sources and from non-masked WIMP search data. The calibrations 
were also used to monitor detector stability and to characterize detector performance. One Si detector 
had a known $^{14}$C contamination, while one Si and one Ge detector showed a poor phonon sensor performance. 
The reported results are thus from 5 Ge and 4 Si ZIPs, chosen before unmasking the WIMP signal region \cite{cdms_si}.

\begin{figure}[h!]
\hspace*{-0.5cm}{\includegraphics[width=.5\textwidth]{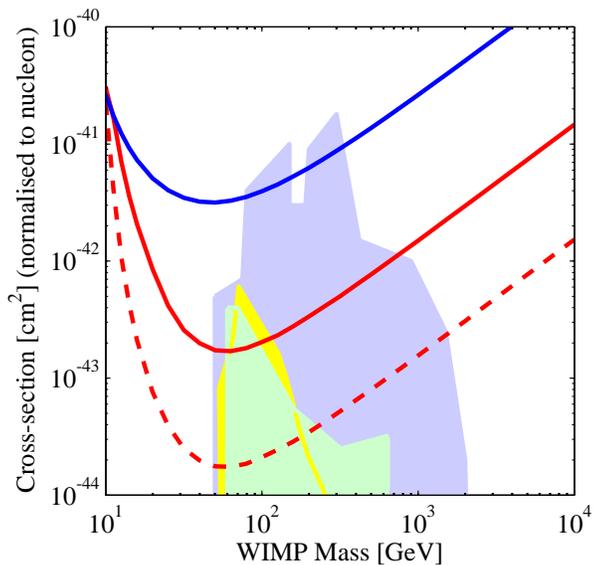}} 
\caption{CDMS results for spin-independent WIMP nucleon cross sections 
versus WIMP mass. Shown are  the Si (blue solid) and Ge (red solid) data
\cite{cdms_si}, as well as the prediction for the 5 towers run at Soudan (red dashed). 
The predicted (filled) regions from supersymmetry are taken from \cite{susy}. Plot 
generated with \cite{dmtools}.} 
\label{fig1b}
\end{figure}

The selection criteria for signal events are based on quantities from the phonon and ionization pulses, 
and are described in detail in \cite{cdms_prd}. To identify surface electron recoils, 5 different 
analyses were performed, the rejection criteria being developed by using events with low ionization yield in the 
$^{133}$Ba data sets. The surface event analysis methods are described in detail in \cite{wang06}.
In the simplest analysis technique, the risetime of the largest phonon pulse, along with its time delay  
with respect to the ionization signal are used to form a timing parameter (shown in 
Fig.~\ref{fig1a}). Since surface events show a smaller timing parameter than most nuclear recoils from the 
$^{252}$Cf source, WIMP-induced nuclear recoils are required to exceed a minimum value for this parameter. 

After un-masking the data, one candidate nuclear recoil event at 10.5\,keV was observed in Ge, while no 
events were seen in the Si data \cite{cdms_si}. 
Although the candidate event occurred in a Ge detector during a period of inefficient ionization collection, 
the result is consistent with the expected background from surface events. 
The derived upper limit on spin-independent WIMP-nucleon cross sections is 1.6$\times$10$^{-43}$cm$^2$ from  
Ge, and 3.4$\times$10$^{-42}$cm$^2$  from Si, both at the 90\%CL and 
at a WIMP mass of 60\,GeV/c$^2$ \cite{cdms_si}. Figure \ref{fig1b} shows the spin-independent 
limits as a function of the WIMP mass, along with theoretical expectations from supersymmetric models. 

\begin{figure}[h!]
\hspace*{-0.5cm}{\includegraphics[width=.5\textwidth]{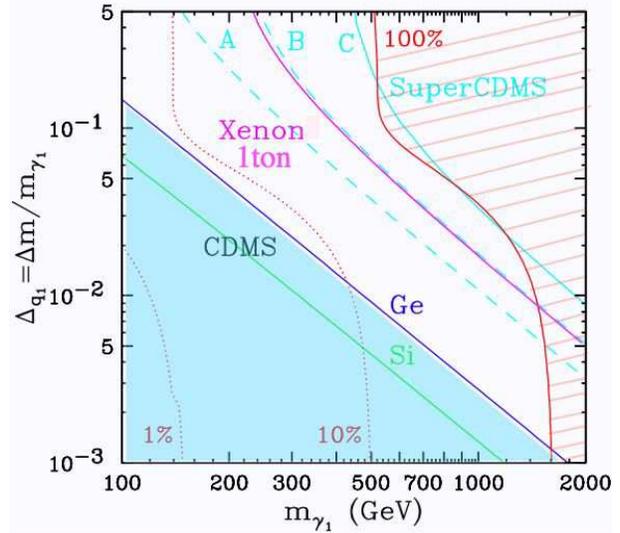}} 
\caption{$\Delta_{q_{(1)}}$ (see text) versus the LKP mass. The CDMS Ge (Si) data excludes 
the region at the left of the line labeled 'Ge' ('Si'). SuperCDMS can probe the parameter space 
at the left of the curves labeled 'A' (25\,kg), 'B' (150\,kg) and 'C' (1\,ton), while XENON-1t will 
probe the space below the solid magenta curve. Also shown are the curves where 1\%, 10\% and 100\% of the 
dark matter is made of LKPs. The hashed region above the 100\% curve is excluded by WMAP results \cite{BKM}.
} 
\label{fig3}
\end{figure}

In theories with flat universal extra dimensions (UED) \cite{UED}, the best studied dark matter candidate is $\gamma_{(1)}$, 
the first Kaluza-Klein (KK) mode of the hypercharge gauge boson. 
Its thermal relic density has been calculated in \cite{ST03}  and \cite{KM}, including 
all the relevant coannihilation processes. It is compatible with WMAP results for a wide range of 
LPK masses, and depends for instance on the masses of the first level KK quarks \cite{KM}. 
In considering the direct detection of $\gamma_{(1)}$ through elastic scattering 
off nuclei, one starts, as in the case of the neutralino, at the quark level. The interaction takes place through 
KK-quark and Higgs exchange, and in the extreme non relativistic limit divides into a spin-independent and a 
spin-dependent part. While q$_{(1)}$ exchange contributes to both scalar and spin-dependent 
couplings, Higgs exchange contributes only to the former one. Since the contributions to the scalar 
interaction interfere constructively, there is a lower bound on both cross sections \cite{CFM}.
Following \cite{CFM} and \cite{ST02}, the  masses 
of the first level quarks (all degenerate), q$_{(1)}$ are taken as a free parameter, and the ratio 
$\Delta_{q_{(1)}}$=$(m_{q_{(1)}}-m_{\gamma_{(1)}})/m_{\gamma_{(1)}}$, which quantifies the mass splitting 
between $\gamma_{(1)}$ and the KK-quarks, is varied from 10$^{-3}$ to 0.5. 
The reach of CDMS is then calculated and shown in the 2-dimensional plane of $\Delta_{q(1)}$  
versus the LKP mass in Figure~\ref{fig3} \cite{BKM}. 
The region excluded by the CDMS  Ge (Si) data is at the left of the curve labeled 'Ge' ('Si). 
The expected sensitivity  
of future experiments such as SuperCDMS \cite{schnee05} and XENON1t \cite{xenon} are also shown.  
The region above the red curve labeled 100\% is excluded by WMAP data, 100\% corresponding 
to $\Omega_{LKP}$ = 0.27 \cite{BKM}.

At present, 5 ZIP towers are  installed in the Soudan cryostat, for a total 
of 4.5\,kg of Ge and 1\,kg of Si. The 30 detectors have successfully been cooled down to base 
temperature and their performance is being checked with gamma calibration events. 
The WIMP search run scheduled to start in fall 2006  is expected to yield a 
factor of $\sim$10 increase in sensitivity (dashed curve Fig.~\ref{fig1b}). 
The  proposed SuperCDMS program \cite{schnee05} is based on 640\,g ZIP detectors with total masses of 25\,kg, 150\,kg 
and 1\,ton. With a target sensitivity of  $\sim$1$\times$10$^{-45}$cm$^2$,  already the 25\,kg phase would be 
complementary to LHC/ILC in the search for new physics at the weak scale.
As shown in Fig.~\ref{fig3}, SuperCDMS-1t  could test the entire region allowed by the WMAP data for Kaluza-Klein 
dark matter in UED \cite{BKM}.


\begin{thebibliography}{9}
 \bibitem{cdms_prd} D.~Akerib {\it et al.} (CDMS Collaboration), {\it Phys.Rev.} D {\bf 72} (2005) 052009.
 \bibitem{cdms_si} D.~Akerib {\it et al.} (CDMS Collaboration),  Phys. Rev. Lett. 96 (2006) 011302.
 \bibitem{wang06} G.~Wang et al., these proceedings.
  \bibitem{susy} H.~Baer {\it et al.}, {\it JCAP}  {\bf 0309} (2003) 007,
E. Baltz and P. Gondolo,  {\it Phys. Rev.} D {\bf 67} (2003) 063503, 
J. Ellis {\it et al.}, {\it Phys.Rev.} D {\bf 71} (2005) 095007. 
\bibitem{dmtools} R. Gaitskell, V. Mandic, J. Fillipini, http://dmtools.berkeley.edu/limitplots/ 
\bibitem{UED}  T.~Appelquist, H.-C. Cheng and B.A. Dobrescu, Phys. Rev. D {\bf 64} (2001) 035002. 
\bibitem{ST03} G.~Servant and T. Tait, Nucl.Phys. B {\bf 650} (2003) 391-419.
\bibitem{KM}  K.~Kong and K.T. Matchev, JHEP {\bf 0601} (2006) 038.
\bibitem{CFM} H.-C. Cheng, J.L. Feng and K.T. Matchev, Phys.Rev.Lett. {\bf 89} (2002) 211301.
\bibitem{ST02} G.~Servant and T.M.P. Tait, New J.Phys. {\bf 4} (2002) 99.
\bibitem{BKM} L.~Baudis, K.C. Kong and K. Matchev, in preparation.
\bibitem{schnee05} R.W.~Schnee {\it et al.}, astro-ph/0502435, 
  P.L. Brink {\it et al.}, astro-ph/0503583.
\bibitem{xenon} E.~Aprile {\it et al.} (XENON) New Astr. Rev. {\bf 49} (2005) 289.


\end{thebibliography}
\end{document}